\documentclass[aps,apl,twocolumn,superscriptaddress,showpacs]{revtex4-1}
\usepackage{latexsym}
\usepackage[dvips]{graphicx}
\usepackage[dvips]{color}
\usepackage{graphicx}
\usepackage{amsmath}

\begin{document}

\title{SQUIDs based set-up for probing current noise and correlations in three-terminal devices}

\author{A. H. Pfeffer, B. Kaviraj, O. Coupiac and F. Lefloch \\ SPSMS/LaTEQS, UMR-E 9001, CEA-INAC and Universit\'e Joseph Fourier, Grenoble - France}
\email[Corresponding author: ]{francois.lefloch@cea.fr}

\begin{abstract}
We have implemented a new experimental set-up for precise measurements of current fluctuations in three-terminal devices. The system operates at very low temperatures (30 mK) and is equipped with three SQUIDs as low noise current amplifiers. A SQUID input coil is connected to each terminal of a sample allowing the acquisition of time-dependent current everywhere in the circuit. From these traces, we can measure the current mean value, the noise and cross-correlations between different branches of a device. In this paper we present calibration results of noise and cross-correlations obtained using low impedance macroscopic resistors. From these results we can extract the noise level of the set-up and show that there are no intrinsic correlations due to the measurement scheme. We also studied noise and correlations as a function of a DC current and estimated the electronic temperature of various macroscopic resistors.
\end{abstract}


\maketitle

\section{Introduction}
In disordered electrical macroscopic conductors, the amplitude of current flowing in response to an applied voltage is governed by various microscopic mechanisms such as elastic scattering on impurities and inelastic collisions within electrons or between electrons and phonons. The conductance, given by the ratio between the mean value of the current and the applied voltage, averages out all the contributions and is usually well described by the Drude formula. At finite temperature and zero applied voltage, the current fluctuates in time around its mean value because of thermal fluctuations of the electronic state occupation number. The amplitude of fluctuations is given by the fluctuation-dissipation theorem and corresponds to the second moment of current distribution. Through that relation, the spectral density of current-noise is proportional to both the temperature and mean value of conductance. In that sense, thermal noise measurement does not provide any additional information than the conductance. At finite applied voltage, scattering events randomize the flow of electrons and induce, together with the granularity of charge, an additional noise contribution known as 'shot noise'.
However, this contribution vanishes in macroscopic disordered conductors when the events are uncorrelated. It is only when the size of the conductor becomes of the order of phase coherence length of the electronic wave function, that shot noise becomes sizable \cite{Blanter00}. This cross-over corresponds to entering the mesoscopic world where quantum effects become important. Since past twenty years, shot noise measurements have contributed to understanding of electronic transport properties at nanoscale. One of the most emblematic results obtained was the experimental demonstration of existence of fractional charges in two-dimensional electrons gas in fractional quantum Hall regime \cite{dePicciotto97, Saminadayar97}. Another very interesting situation was obtained in hybrid superconducting (S) / normal metal (N) nanostructures where the shot noise was found to be doubled (SN junctions) or very much enhanced (SNS junctions) due to conversion of Cooper pairs into normal electrons at the S/N interface \cite{Jehl00, Lefloch03, Hoffmann04, Lhotel07}. \\
But the interest of fluctuation measurements is not limited to noise. For example, the third moment of current distribution reveals its (non-)Gaussian character whereas noise correlations can probe statistics (fermionic of bosonic) of current carriers like in a Hanbury-Brown Twiss type of experiment \cite{Hanbury56}. Cross-correlations can also be investigated to reveal splitting of Copper pairs in hybrid normal-superconducting nanostructures \cite{Anantram96, Martin96}.\\
However, in order to address experimentally these quantities, new instruments need to be developed. The pioneer work on the third moment of a simple tunnel junction has revealed a major importance of the environment showing that these quantities need to be investigated with great care \cite{Reulet03, Beenakker03}. The same is true for cross-correlations where the signal is much smaller than the noise and where all spurious contributions need to be clearly identified \cite{Wei10, Das12}. \\

\section{Instrumental set-up}
\begin{figure}
\includegraphics[width=8.5cm]{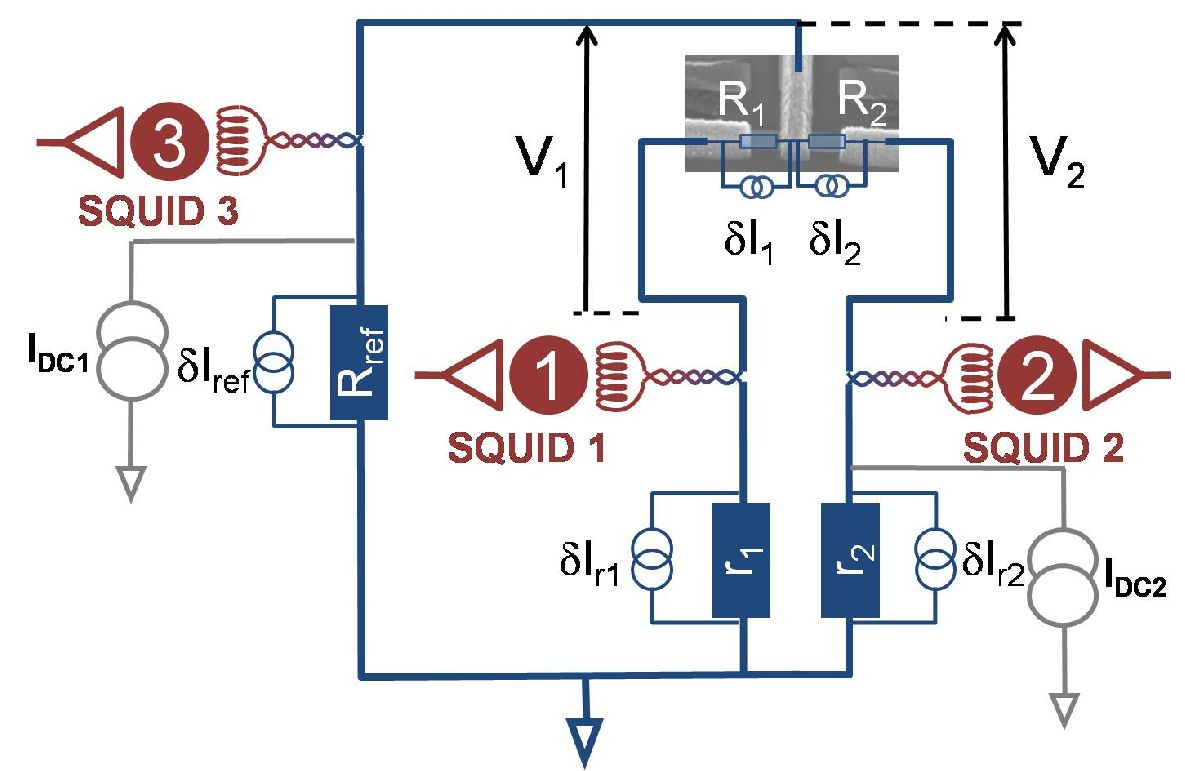}
\caption{Schematic description of the experimental set-up. 
The schematic shows a typical three-terminal nanodevice to be studied with this new instrument.The red circuit's elements are at 4.2 K and the blue ones are thermalized with the mixing chamber of the dilution fridge. For calibration measurements, the sample has been replaced by two macroscopic resistors of $\sim 1 \, \Omega$ resistance whereas the others macroscopic resistors in the circuit have a resistance of $\sim 0.1 \, \Omega$. All the wiring is made from superconducting leads.}
\label{Fig1}
\end{figure}

In our set-up, the mean value and fluctuations of currents in each branch of a sample are directly measured using three SQUIDs as current amplifiers (see figure \ref{Fig1}). The measurements are done at very low temperatures, using a dilution fridge with a base temperature of 30 mK. The SQUID electronics operate in the standard flux lock loop mode which reduces almost completely the spurious feed-back of read-out towards the sample. 
As current fluctuations are measured, the sample needs to be voltage-biased. This is achieved by placing, in parallel to the sample, a macroscopic "surface mount resistor" $R_{ref}$ whose resistance is much lower than the device resistance and that is mounted at low temperature to reduce its noise contribution (see Fig. \ref{Fig1}). 
Then, a low noise current source $I_{DC1}$ is used to bias the system. Two additional resistors $r_1$ and $r_2$ are placed in series with two other branches of the three-terminal device. With a second low noise current source $I_{DC2}$, it is then possible to control the two voltage drops $V_1$ and $V_2$ independently. Additionally, each individual voltage drop can be measured using room temperature low noise differential voltage amplifiers. If needed, the differential resistance can also be measured by sending a small ac-current together with the DC-current.\\
As current noise is generally inversely proportional to the sample resistance, our set-up is well adapted to low impedance samples, typically around $1 \, \Omega$.
This statement is exact for thermal noise and can be somewhat generalized to shot noise for hybrid superconducting nanostructures. Indeed, when superconductivity is involved, it exists a typical voltage scale given by the superconducting gap and the current needed to reach this gap is inversely proportional to the sub-gap impedance of the device. 
In practice, the spectral density of noise can be measured with a set-up noise of $2\sim 3 \, 10^{-24} \,A^2$/Hz = $2\sim 3$ p$A^2$/Hz . This noise floor does not strictly correspond to current fluctuations at the input coil of the SQUIDs but mostly come from the noise of the room temperature electronics and translates into these values using the overall gain of the amplification/conversion chain. The bandwidth is given by the resistance of the overall device (sample in parallel with the reference resistor, $R_{ref}$) coupled to the input coil inductance. Using a $0.1 \,\Omega$ reference resistor and given the SQUIDs input coil inductances ($L \sim 1 \, \mu H$), we obtain a cut-off frequency $f=R_{ref}/(2\pi L) \sim 15$ kHz that limits the bandwidth. Another important advantage of using SQUIDs is the very low 1/f noise that is sizable only below 1 Hz. With these values and considering a reasonable acquisition time of few minutes, we obtain a sensitivity of order of 0.1 p$A^2$/Hz.\\
Figure \ref{Fig1} shows the schematic of the experimental set-up together with the microscope image of a typical three-terminal device \cite{Kaviraj11}. For this sample, a central superconducting electrode emits Cooper pairs towards two spatially separated superconducting collectors through two normal metal nano-bridges of $1.5 \mu m$ length. The sample can be modeled by two non-linear resistors $R_1$ and $R_2$. All the wires connecting different parts of the set-up are superconducting and therefore do not add any additional resistance in the circuit. The three SQUIDs sensors are located at the 4.2 K flange of the dilution refrigerator in the helium liquid bath. All other elements are thermalized with the mixing chamber temperature. 

Any resistive part of the circuit is associated with a current source $\delta I$ in parallel, in accordance with Nyquist representation. In this model, the current flowing in each SQUID is given by :
\begin{equation}
\begin{gathered}
\delta I_{sq}^i=\left(\frac{\widehat{R_i}}{\widehat{R_i}+\widehat{R_{jk}}} \right)\widehat{\delta I_i} - \left(\frac{\widehat{R_{jk}}}{\widehat{R_i}+\widehat{R_{jk}}} \right) \left(\widehat{\delta I_j}+ \widehat{\delta I_k} \right)
\label{EquSqCurr}
\end{gathered}
\end{equation} 
where $\widehat{R_i}$ is the sum of the resistances in branch $i$, $\widehat{\delta I_i}$ the weighted current contribution of all the elements in branch $i$ and $\widehat{R_{jk}}$ the equivalent resistance of $\widehat{R_j}$ in parallel with $\widehat{R_k}$. For instance, 
\begin{equation}
\begin{gathered}
\widehat{R_1}=R_1+r_1 \hspace{1cm} \widehat{R_3}=R_{ref}\\
\widehat{\delta I_1}=\frac{R_1}{\widehat{R_1}}\delta I_1+\frac{r_1}{\widehat{R_1}}\delta I_{r1} \hspace{1cm} \widehat{\delta I_3}=\delta I_{ref}.
\end{gathered}
\end{equation}   

From the three fluctuating SQUID currents $\delta I_{sq}^i$, we can perform three auto-correlations $AC_i\equiv \delta I_{sq}^i \delta I_{sq}^i$ and three cross-correlations $XC_{ij}\equiv \delta I_{sq}^i \delta I_{sq}^j$ non-independent measurements given by:
\begin{equation}
\begin{gathered}
AC_i=<FFT^*(\delta I_{sq}^i) FFT(\delta I_{sq}^i)>\\
XC_{ij}=<FFT^*(\delta I_{sq}^i) FFT(\delta I_{sq}^j)>
\label{EquACXC}
\end{gathered}
\end{equation}
where $FFT$ stands for the $Fast\, Fourier\, Transform$, $FFT^*$ its complex conjugate and $<...>$ the rms average at the spectrum analyzer. At this stage, the importance of voltage biasing can be seen. Indeed, in the case of a current biasing scheme, the resistance value of the reference resistor would be much larger than all the other resistances. Applying this condition to the set of equations, it becomes obvious that the correlations are always negative as the result of current conservation applied to current fluctuations : $(\delta I_{sq}^1+\delta_{sq}^2)^2=0$. \\
In principle, the 6 measurements are related to the noise and cross-correlations of each element of the circuit through a 6x15 matrix ! This can be very much simplified by considering that any cross-correlated noise involving at least one macroscopic resistor is zero.
Therefore, the system of equations reduces to :
\begin{equation}
\begin{bmatrix}
AC_1\\AC_2\\AC_3\\XC_{12}\\XC_{13}\\XC_{23}
\end{bmatrix}
= M_{tot}
\begin{bmatrix}
S_1\\S_2\\S_{12}
\end{bmatrix}
+ N_{tot}
\begin{bmatrix}
S_{r1}\\S_{r2}\\S_{ref}
\end{bmatrix}
\label{MatrixTot}
\end{equation}
where $S_{ref}, S_{r1}$ and $S_{r2}$ are the thermal noise of the three macroscopic resistors $R_{ref}, r_1$ and $r_2$. 
The three quantities of interest are the noise in each branch of the device i.e $S_1\equiv \delta I_1\delta I_1$ and $S_2\equiv \delta I_2\delta I_2$ and the cross-correlated noise $S_{12}\equiv \delta I_1\delta I_2$. Therefore only three measurements are necessary and we usually choose $AC_1, AC_2$ and $XC_{12}$ that are related to the physical quantities in the following way:
    
\begin{equation}
\begin{bmatrix}
AC_1\\AC_2\\XC_{12}
\end{bmatrix}
= M
\begin{bmatrix}
S_1\\S_2\\S_{12}
\end{bmatrix}
+ N
\begin{bmatrix}
S_{r1}\\S_{r2}\\S_{ref}
\end{bmatrix}
\label{Matrix}
\end{equation}

Following equation \ref{EquSqCurr}, the matrix elements are just given by the values of various resistances. Note that in the case of a sample with non-linear response, $R_1$ and $R_2$ are differential resistances.       
As an example, the two 3x3 matrices $M$ and $N$, for $R_1=R_2=1.0\, \Omega$ and $R_{ref}=r_1=r_2=0.1\, \Omega$, are :
{\footnotesize
\begin{equation}
\begin{gathered}
M = 
\begin{bmatrix}
0.7072&0.0049&-0.1179\\
0.0049&0.7072&-0.1179\\
-0.0589&-0.0589&0.7105\\
\end{bmatrix}
\\N = 
\begin{bmatrix}
0.007072&0.000049&0.0059\\
0.000049&0.007072&0.0059\\
-0.000589&-0.000589&0.0059\\
\end{bmatrix}
\end{gathered}
\end{equation}
}
From this example, we see that the $AC_1$ ($AC_2$) is mostly due to the noise $S_1$ ($S_2$). Focusing on $XC_{12}$, it reads :

\begin{align}
	XC_{12}= & -\mid M_{31} \mid S_1-\mid M_{32} \mid S_2+\mid M_{33} \mid S_{12} \notag \\
	& -\mid N_{31} \mid S_{r1}-\mid N_{32} \mid S_{r2}+\mid N_{33} \mid S_{ref}
\label{XC12}	
\end{align}
where $\mid M_{ij} \mid$ refers to the absolute value of the matrix element $M_{ij}$. This notation shows that the sign of $XC_{12}$ is not necessary that of the crossed correlated noise $S_{12}$ and that the contribution $S_{ref}$ is always positive. 
There are different ways to extract $S_{12}$ from $XC_{12}$. First, if all the contributions other than $S_{12}$ are known, the crossed correlated noise of a sample is known simply by removing those contributions from  $XC_{12}$ and divide by $M_{33}$. Usually, $S_1$ and $S_2$ are not known. In that case, the matrix $M$ in equation \ref{Matrix} can be inverted and both the spectral densities of noise $S_1$ and $S_2$, and the cross-correlations $S_{12}$ can be obtained from measurements. 

\section{Calibration}
\begin{figure}
\includegraphics[width=8.cm]{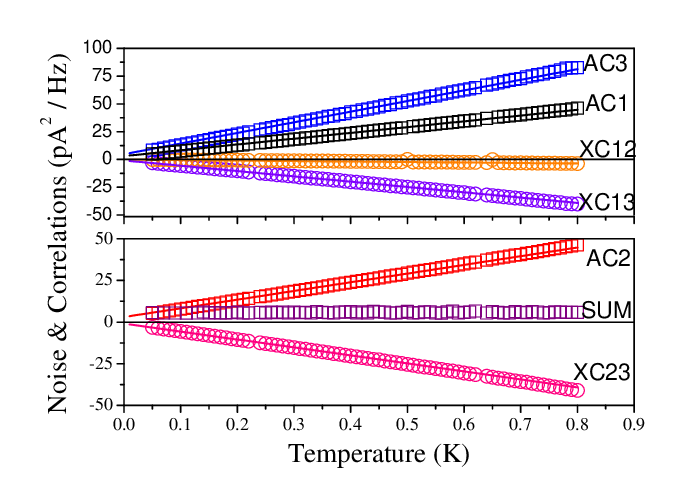}
\caption{Noise $AC_i$ and correlations $XC_{ij}$ of the three SQUIDs in response to the circuit depicted in figure \ref{Fig1} as a function of the mixing chamber temperature with zero DC current. The "SUM" accounts for the current conservation law and the solid lines are adjustments of the data using equations \ref{MatrixTot} and the appropriate offsets (see text).}
\label{Fig2}
\end{figure}

In order to test the performance of our instrument, we have measured noise and correlations of a test sample made from two macroscopic resistors of $1 \, \Omega$ resistance at room temperature. For this test, these two macroscopic resistors are placed on the sample holder whereas the biasing resistors $R_{ref}$, $r_1$ and $r_2$ are anchored to the mixing chamber. We then expect the current noise in each branch to be given by the thermal noise contribution of each resistive element of the circuit assuming no intrinsic cross-correlated noise between them as they are all macroscopic. The overall results of the six measurements are shown in figure \ref{Fig2} as a function of the temperature. In this plot, the "SUM" is $\left (\Sigma \delta I _{sqi}\right )^2=\Sigma AC_i+2\Sigma XC_{ij}$ accounting for the current conservation law. The precise measurement of the resistances at low temperature gives: $R_1=R_2=0.88\, \Omega$ and $r_1=r_2=R_{ref}=0.088\, \Omega$. 

The solid line in figure \ref{Fig2} show the noise and correlations estimates given by equation \ref{MatrixTot} considering only thermal noise and setting $S_{12}=0$.
The only adjusting parameter is then the set-up noise contribution that can be read extrapolating the data at zero temperature. We obtain a set-up noise level of the order of 2-3 p$A^2$/Hz for the three auto-correlation $AC_i$ measurements. The set-up noise level correspond to a flux noise of few $\mu \Phi_0/\sqrt{Hz}$ in a SQUID, which is close to the best noise level given by the manufacturer considering the overall experimental set-up and its connections to room temperature instruments. Note that these measurements have been performed with one DC-current source connected to the circuit but with zero current at its output. Therefore, our results include any additional noise coming from this external source. Regarding the cross-correlation between the SQUID currents $XC_{ij}$, there are no fitting parameters as we have fixed the intrinsic correlations $S_{ij}$ to be zero. The agreement with the data is very good so we conclude that there is no measurable intrinsic correlations between the three SQUID output signals. \\

\begin{figure}
\includegraphics[width=8.cm]{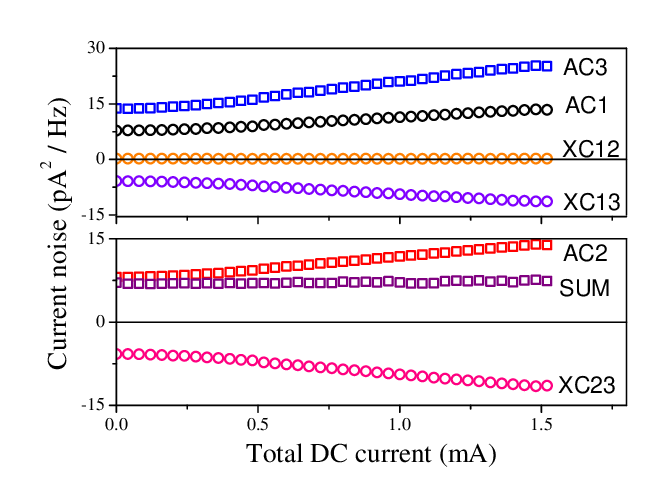}
\caption{Noise $AC_i$ and correlations $XC_{ij}$ of the three SQUIDs as a function of the total DC current $I_{DC1}$ applied on the reference resistor side at $100 mK$.}
\label{Fig3}
\end{figure}

In a second test, we have measured the noise and correlations as a function of a DC-current at fixed temperature. We used $I_{DC1}$ for this measurement and ramped the current up to 1.5 mA at fixed temperatures of 100 mK, 300 mK and 500 mK. The overall measurements are shown in figure \ref{Fig3} for a base temperature of 100 mK. As all the resistive elements in the circuits are macroscopic resistors with temperature independent resistances, the shot noise is zero and the noise response should stay constant and given by the thermal noise. However, we clearly see an increase of the noise and correlation responses.  This increase corresponds to an elevation of the electronic temperature of the resistors due to Joule heating whereas the temperature of the mixing chamber was kept regulated at 100 mK. From these results, we can estimate the effective temperature of the resistors. 
However, at a given value of the DC current, the effective temperatures of the various resistors are not equal.  
There are two reasons for that. First, because the biasing resistors $r_1, r_2$ and $R_{ref}$ are anchored to the mixing chamber their coupling to the cold bath is better than for the $R_1$ and $R_2$ resistors that are placed and connected similarly to what a real sample will be i.e on a sample holder.  Second, for a given DC current $Idc1$, the current flowing through $R_{ref}$ is different from the current though the other resistors. 
Thanks to the symmetry of the circuitry, we can consider that $T_{R1}=T_{R2}\equiv T_R$ and $T_{r1}=T_{r2}\equiv T_r$, both sets being different from $T_{ref}$ where $T_{R}$ is the effective temperature of a resistor $R$. 

\begin{figure}
\includegraphics[width=9.5cm]{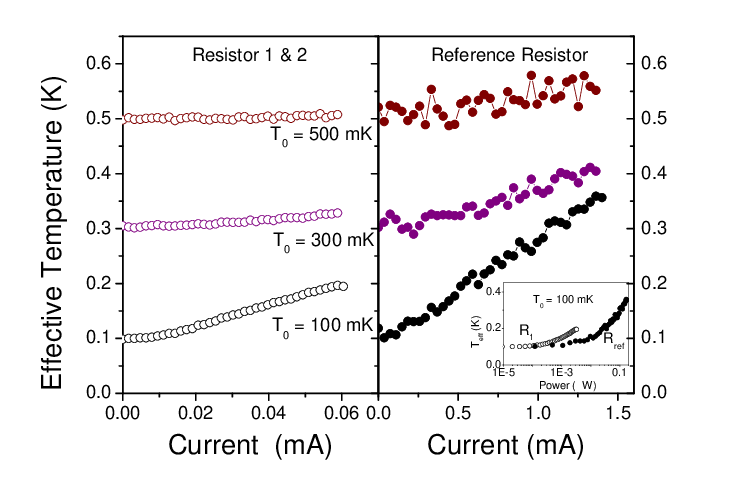}
\caption{Electronic temperature of the reference resistor $R_{ref}$ and of the sample resistors $R_1$ and $R_2$ as a function of the dc current flowing through each of them for three different regulated temperatures. The inset shows the effective temperatures at a base temperature of 100 mK as a function of the dissipated power.}
\label{Fig4}
\end{figure}

In order to estimate the effective temperatures in the simplest way, we shall be using only the $AC_1$ and $AC_3$ measurements (equation \ref{MatrixTot}). 
Considering only thermal noise, the two equations become :

\begin{equation}
\begin{gathered}
AC_1=\alpha_1 T_R+\alpha_2 T_r+ \alpha_3 T_{ref} + AC_1^0 \\
AC_3=\beta_1 T_R+\beta_2 T_r+ \beta_3 T_{ref} + AC_3^0
\label{EqTeff}
\end{gathered}
\end{equation}
where $\alpha_1=4 k_B\times (M_{tot11}/R_1+M_{tot12}/R_2)$ for instance and $AC_1^0$ and $AC_3^0$ the set-up noise of the SQUIDs 1 and 3 obtained from the measurements as a function of the temperature.
We can simplify these equations by neglecting the noise contribution of the resistances $r_1$ and $r_2$. This assumption is justified, first because $\alpha_2 (\beta_2)$ is ten times smaller than $\alpha_1 (\beta_1)$ and second because the current through $r_1$ and $r_2$ is much smaller than though $R_{ref}$ \cite{Note}. The effective temperatures $T_R$ and $T_{ref}$ are then extracted from equations \ref{EqTeff}.     
The results are shown in figure \ref{Fig4} for various base temperatures. We do see that the temperature increase is weaker as the base temperature is raised and is almost absent above $500 mK$. This behavior is due to the strengthening of the electron-phonon coupling as the temperature is raised. The temperature increase is clearly more pronouced at the reference resistor than for the sample resistors and reaches $\Delta T_{ref} \sim$ 250 mK and $\Delta T_R \sim$ 100 mK  at the lowest temperature investigated (i.e 100 mK). In the inset, we have plotted the effective temperatures as a function of the power $R\,I^2$ that is dissipated in each resistor. The difference traces for $T_R$ and $T_{ref}$ reflects the fact that the coupling to the mixing chamber of the biasing resistors or the sample resistors are different. \\
The increase of effective temperature in the biasing resistors translates directly into an increase of their noise contribution in the measured quantities (see equation \ref{Matrix}) and may affect the estimate of the quantities of interest. However, because the corresponding matrix elements are small, this additional contribution stays very moderate but needs to be considered in real samples especially for the $XC_{12}$ measurement.\\

\section{Conclusion}
In conclusion, we have described the performances of an experimental instrument implemented to measure current noise and correlations of three terminal nanostructures. This set-up is particularly well adapted to low impedance samples and uses three SQUIDs as current amplifiers. We have calibrated the overall response of the system using macroscopic resistors in place of a real sample. We have shown that despite the complexity of the instrument, we can reach very low intrinsic noise floor and that there is no significant intrinsic noise correlations between the SQUIDs outputs. Finally, the increase of the effective temperature of the various resistive elements of the circuit has been measured as a function of a DC-current.\\

\section{Acknowledgment }
This work has been partially financed by the French Research National Agency (ANR) through two main contracts (ACI-JC 2003 and ANR-PNANO 2008). We acknowledge the RTRA Nanscience foundation for its support to the PhD grant of A. Pfeffer.


\begin{thebibliography}{10}
\bibitem{Note} We have checked that more elaborated procedures, taking into account the noise of the resistors $r_1$ and $r_2$, give the same result.
\end{thebibliography}

\end{document}